\documentclass[prc,showpacs,twocolumn]{revtex4}
\usepackage{graphicx}

\def\pt{$p_T$}
\def\bg{background}
\def\dis{distribution}
\def\st{statistical}
\def\fq{$F_q$}

\begin{document}

\title{
Suppression of statistical background in the event structure of away-side $\Delta\phi$ distribution}

\author{Charles B.\ Chiu$^1$ and  Rudolph C. Hwa$^{2}$}
\affiliation
  {$^1$Center for Particle Physics and Department of Physics, University of Texas at Austin, Austin, TX 78712, USA\\
$^2$Institute of Theoretical Science and Department of
Physics, University of Oregon, Eugene, OR 97403-5203, USA}

\begin{abstract} 
An approach is proposed to analyze the azimuthal distribution of particles produced on the away side in heavy-ion collisions without background subtraction. Measures in terms of factorial moments are suggested that can suppress the statistical background, while giving clear distinction between one-jet and two-jet event structures on the away side. It is also possible to map the position and strength of the recoil jet to suitably chosen asymmetry moments.
\end{abstract}

\maketitle
	An important area of investigation in relativistic heavy-ion collisions is the  azimuthal angular distribution of hadrons produced on the opposite side of particles triggered at high \pt\ \cite{ja,hb,dm}. That distribution reveals the properties of the medium effect on a hard recoil parton traversing the dense system formed by a central collision. Several possibilities have been suggested for what may happen to the hard parton and what signatures may be registered on the away side \cite{cst,kmw,id}. Our concern in this paper is not so much on the nature of the possible signatures, but on how best to extract them from a sea of noisy background that is inherent in every nuclear collision. The conventional approach is to subtract the \bg.  That is accomplished by first summing over all events subject to specific kinematical cuts. It is meaningless and impossible to make \bg\ subtraction event by event. However, summing over all events is a process in analysis that often degrades the signal, since the recoil parton may be absorbed by the medium, or may emerge on the other side as a reduced minijet, or may generate a shock wave. To enhance the signature what is needed is a measure that filters out the statistical fluctuation in every event to the extent possible, and then let that measure be summed over all events. The aim of this paper is to propose such a measure and to demonstrate its effectiveness in the framework of a simple model that simulates events with jets in the midst of a large statistical \bg.

	The measure that we propose is factorial moments. Before going into the details, let us first give some historical \bg. The factorial moments were first used in the study of intermittency, which quantifies the self-similar behavior of branching processes in multiparticle production \cite{bp,ddk}. It has also been suggested for critical behavior in heavy-ion collisions \cite{bh}. Although intermittency has been found in elementary collisions, its study in nuclear collisions has been plagued by a number of difficulties, among which is the problem of \st\ fluctuations in large systems despite the theoretical virtues of the factorial moments. There are several differences between those intermittency studies and our problem here. First, we do not analyze fluctuations in rapidity, and do not consider a wide range of bin size to search for fractal behavior. Second, we consider only a subset of rarer events selected by high \pt\ triggers, and examine the away-side $\Delta\phi$ distribution that consists of far lower multiplicities of particles. Third, in intermittency studies the dynamical fluctuation is convoluted with statistical fluatuation, whereas our jet signal on the away side is additive relative to the \bg. Having made these remarks to dissociate our work from whatever vestige there may be from the past, we proceed now to a description of our measure from the basics.
	
		The usual $\Delta\phi$ variable is defined relative to the trigger momentum in the transverse plane. For the convenience of our analysis on the away side, let us define
$\varphi=\Delta\phi-\pi$. 
Consider an interval $I$ around $\varphi=0$, which, for definiteness, may be taken to be from $-1.5$ to +1.5, although the suitable range is an experimental decision. Let $I$ be divided into $M$ equal bins so that the bin size is 
$\delta=I/M$. 
In an event let $n_j$ denote the number of particles in $j$th bin. Define the factorial moment (FM) by
\begin{eqnarray}
f_q={1\over M}\sum_{j=1}^M n_j(n_j-1)\cdots(n_j-q+1)  \label{1}
\end{eqnarray}
and the normalized factorial moment (NFM) by
\begin{eqnarray}
F_q=f_q/f_1^q,   \label{2}
\end{eqnarray}
for each event.  It is important to recognize that Eq.\ (1) is used to determine the FM  event by event, and the summation there may be regarded as the horizontal average.

	Let us now consider the hypothetical case where the fluctuation of $n_j$ from bin to bin can be described by a Poisson \dis\ as an ideal representation of the \bg\ 
	\begin{eqnarray}
P_{\bar n}(n)={\bar n^n\over n!}e^{-\bar n} ,   \label{3}
\end{eqnarray}
where $n$ is the bin multiplicity that fluctuates and $\bar n$ is
the average that  depends on $\delta$ and total event multiplicity $N$. Then Eq.\ (\ref{1}) can be written as
\begin{eqnarray}
f_q=\sum_{n=q}^N {n!\over (n-q)!} P_{\bar n}(n) .   \label{4}
\end{eqnarray}
If $N$ is large compared to $\bar n$, it can be approximated by infinity in Eq.\ (\ref{4}); it then follows from  Eq.\ (\ref{3}) that $f_q=\bar n^q$, so we have
$F_q^{({\rm stat})} = 1$   for any $\delta$ in the ideal \st\ case. 
When $N$ is finite but large, the corresponding $F_q^{({\rm stat})}$ should be slightly less than 1.
Furthermore, it has been pointed out to us that $F_q$ as defined in Eq.\ (\ref{2}) is independent of detector efficiency \cite{at}. 

In reality, $F_q^{({\rm stat})}$ is not exactly 1. We expect $F_q$ to fluctuate from event to event. Let us define in general the event-averaged NFM
\begin{eqnarray}
\left<F_q\right>(\delta)={1\over {\cal N}_{\rm evt}}\sum_{i=1}^{ {\cal N}_{\rm evt}} F_q^{(i)}(\delta, N_i),  \label{5}
\end{eqnarray}
where $N_i$ is the multiplicity of the $i$th event and $ {\cal N}_{\rm evt}$ is the total number of events. The central question is whether $\left<F_q\right>$ stays approximately at 1 for pure \bg\ in the real data and becomes larger than 1 significantly enough, when jets exist, so that $\left<F_q\right>$ can be used  as an effective measure to distinguish the signal from the noise. We stress that the (vertical) average over all events in Eq.\ (\ref{5}) is performed after the horizontal average is done in Eq.\ (\ref{1}), not the other way around. Thus if the answer to the above central question is in the affirmative, then one may regard $F_q$ as effectively suppressing the \st\ \bg\ event-by-event. Since it is impossible to make \bg\ subtraction event-by-event, to suppress the \st\ contribution to a suitably chosen measure seems to be the best that one can do, and it should be done at the level of each event.

	To test the effectiveness of the measure suggested above, let us consider a simple model to simulate   the $\varphi$ distributions of events with jets in the presence of significant \bg. Let $N_i$ particles be distributed randomly in the interval $I$ with $-1.5<\varphi<1.5$. They will be regarded as the \bg\ of the $i$th event. For now, we limit ourselves to the simplest case of $N_i$ being a constant, but later generalize $N_i$ to be Gaussian distributed. For a 1-jet event, we add a cluster of particles of multiplicity, $m=5$,  bunched in a small  interval in $\varphi$ of extent $\epsilon=0.04$. The cluster is randomly located  in the interval $-1<\varphi<+1$.  For a 2-jet event (to mimic a Mach cone structure) we put in two clusters of 5 particles each. We label the three cases by bg, 1j and 2j, respectively, where bg denotes background only without jets. In Fig.\ 1 we show the distributions of (a) $F_2$ and (b) $F_3$ for the three cases, for $N_i=60, M=30$ and ${\cal N}_{\rm evt}=2000$. Although the \dis s have significant overlap, their average values $\left< F_q\right>$ are distinctly separated, as shown in Fig.\ 2, for various bin numbers $M$. In (a) the average values $\left< F_q\right>$ for the background are all less than, but close to, 1. In (b) for the case including 1-jet and  in (c) for the bg+2-jet case,  $\left< F_q\right>$ are distinctly higher than 1, especially for $q>2$. 
	These results represent the first indication that the FM method is useful, and that the suppression of the \st\ fluctuation is realized upon event averaging without explicit subtraction of the background. That is, with $\left< F_q\right>$ being around 1 for background only, any significant enhancement of $\left<F_q\right>$ above 1 would indicate the presence of non-statistical signal.

 \begin{figure}[tbph]
\hspace*{-0.67cm}
\includegraphics[width=0.55\textwidth]{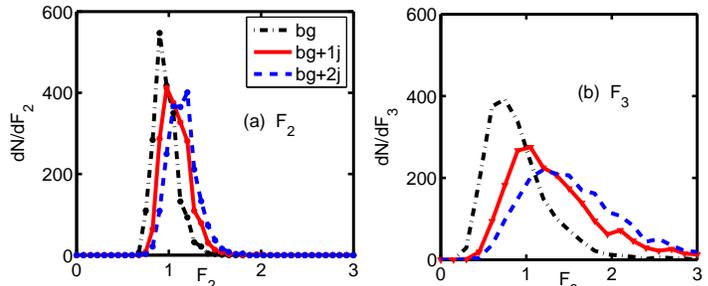}
\vspace*{-1cm}
\caption{Event-by-event distributions of $F_q$ for (a) $q=2$ and (b) $q=3$ for 2000 events with 30 bins.}
\end{figure}
\vspace*{-.7cm}
\begin{figure}[tbph]
\hspace*{-0.7cm}
\includegraphics[width=0.57\textwidth]{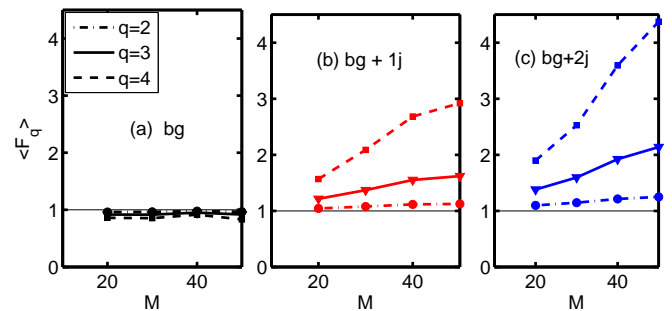}
\vspace*{-0.8cm}
\caption{Event-averaged $\left<F_q\right>$ vs bin number $M$ for (a) bg only, (b) bg+1j, and (c) bg+2j. Circles, triangles and squares are for $q=2,3,4$, respectively.}
\end{figure}

 	In a more realistic situation  elliptic flow introduces a $\varphi$ dependence to the \bg. We have accordingly modified  the statistical \bg\  by a factor $1+2v_2{\rm cos}2\varphi$.  We find that Fig.  2 remains essentially the same for $v_2$ increasing up to 0.1, and that the 1j and 2j cases can be readily distinguished from the \bg\ with flow. 
 The reason is that the bin-to-bin variation of the $2v_2\cos 2\varphi$ factor is small compared to the variation due to a jet in any event, and the FMs are able to distinguish the two types of variation event-by-event. 
 
 	In the test cases studied above we have taken the background to consist of only 60 particles, distributed over an interval of $I=3$, corresponding to $dN/d\varphi|_{bg}\approx 20$. 
	We use $dN/d\varphi$ to denote the event-averaged azimuthal distribution, as is conventionally done, although in our simulation here the same notation is used for the distribution in any event, for brevity's sake.
	In some experimental data the lower bound of $p_T^{\rm assoc}$ has been set as low as 0.15 GeV/c \cite{ja}, for which the background height is as high as $dN/d\varphi|_{bg}\approx 200$. In such cases the signal of jets on the away side is at the 1\% level; it is therefore in the presence of statistical bin-to-bin fluctuations that can be significantly larger, and our method cannot be expected to be effective. We have found that for $dN/d\varphi|_{bg}<50$,  the 1j and 2j signals quantified by $\left<F_q\right>$ can be distinguished from the background.

	A disadvantage in working with \fq\ is that one no longer sees visually the peaks in $\Delta\phi$ \dis\ associated with jets. Thus one gets a quantitative measure of the peaks  at the expense of losing information about the locations of the peaks. However, the loss can be reduced if we elevate the level of horizontal analysis of the FM by being more specific about the spatial regions in $\varphi$. To that end let us define
	\begin{eqnarray}
f_q^{\pm}={1\over M_{\pm}}\sum_{j\in R_{\pm}} n_j(n_j-1)\cdots(n_j-q+1),  \label{6}
\end{eqnarray}
where $R_{\pm}$ stands for the region where $\varphi$ is $\ ^>_<\  0$, and $M_{\pm}=M/2$ is the number of bins in $R_{\pm}$. Thus we have $f_q=(f_q^++f_q^-)/2$ for each event. We further define the NFM, as in Eq.\ (\ref{2}), by 
\begin{eqnarray}
F_q^{\pm}=f_q^\pm/(f_1)^q,  \label{7}
\end{eqnarray}
which implies $F_q = (F_q^{+}+F_q^{-})/2$  for every event. 

	Since event averaging is likely to erase the distinction between $F_q^{+}$ and $F_q^{-}$ if a jet fluctuates between being on the $+$ and $-$ sides, it is useful to consider the difference moments  $D_q=|F_q^{+}-F_q^{-}|$ and the sum $S_q=F_q^+ + F_q^-$. To amplify the effect of the fluctuations let us define the vertical average of the $p$th moment of $D_q$ by
\begin{equation}
\left<D_q^p\right>(\delta)={1\over  {\cal N}_{\rm evt}}\sum_{i=1}^{ {\cal N}_{\rm evt}} {D_q^{(i)}}^p(\delta, N_i),   \label{8}
\end{equation}
and similarly of $S_q$.  A plot of $\left<D_q^p\right>(\delta)$ versus $\left<S_q^p\right>(\delta)$ for  the three cases of bg, 1j and 2j for various values of $\delta$ can reveal some characteristics of interest. In our simulation we place the two jets always on the opposite sides of $\varphi=0$ in order to contrast the 2j from the 1j events.
In Fig.\ 3 we show an array of such plots for $p,q=2,3,4$. In all, the background points are clustered together, while the 1j and 2j points fan out from the origin, mostly in straight lines. Clearly, two jets on two sides of $\varphi=0$ reduce $D_q$, resulting in lower $\left<D_q^p\right>$ compared to the same from 1j.
Similar plots of the experimental data would indicate first of all whether the $\left<D_q^p\right>$ and $\left<S_q^p\right>$ points of the background  cluster together at the lower-left corner, and then secondly  whether the points arising from 1j and 2j are well separated from those from just the background.

\begin{figure}[tbph]
\hspace*{-.5cm}
\includegraphics[width=0.55\textwidth]{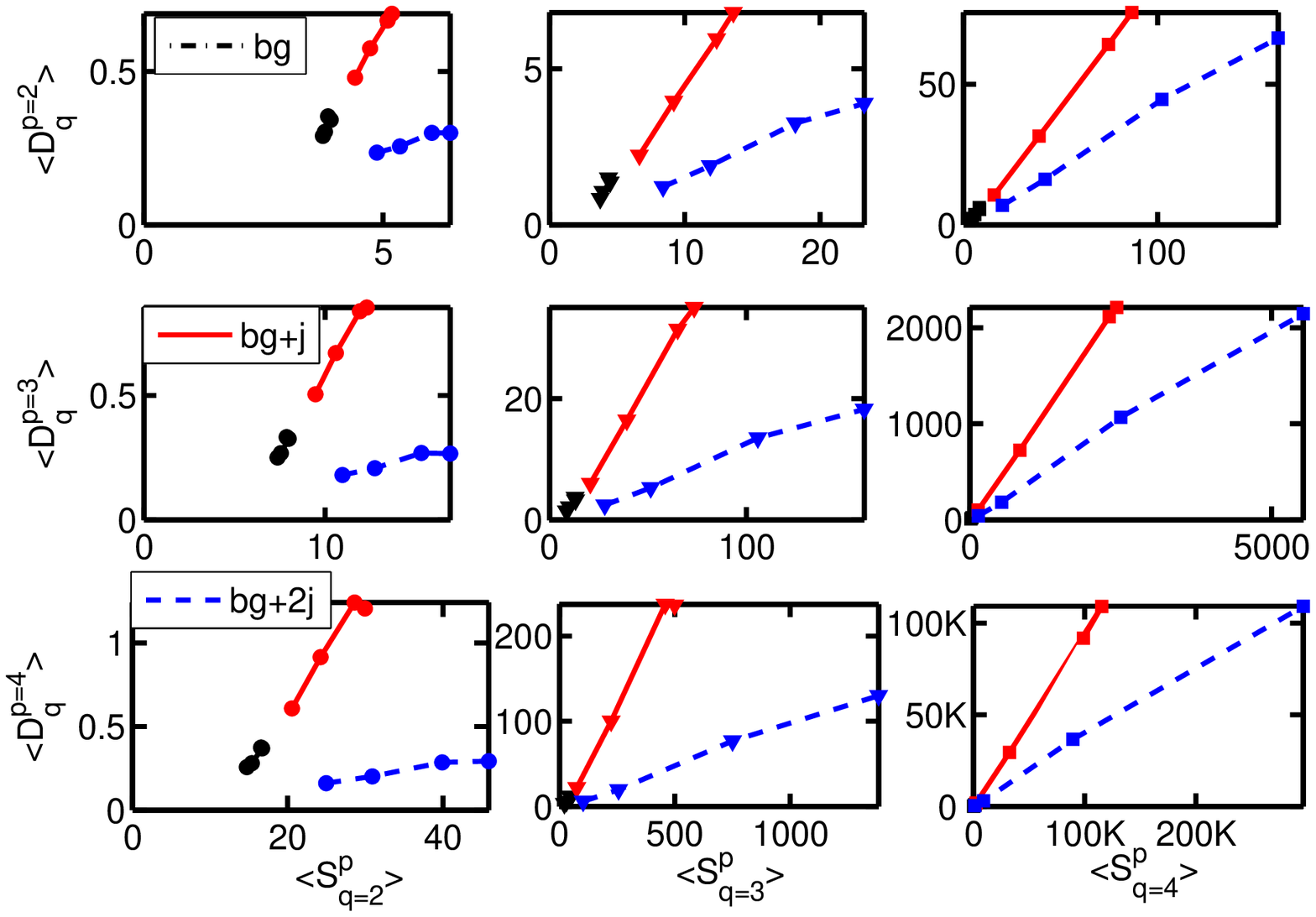}
\vspace*{-.8cm}
\caption{ $\left<D_q^p\right>$ vs $\left<S_q^p\right>$ for $M=20,30,40,50$.}
\end{figure}

\begin{figure}[tbph]
\hspace*{-0.5cm}
\includegraphics[width=0.57\textwidth]{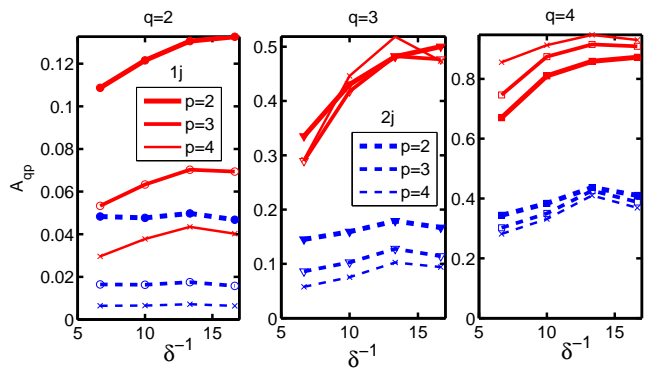}
\vspace*{-.8cm}
\caption{$A_{qp}$ vs $\delta^{-1}$ for (a) $q=2$, (b) $q=3$, and (c) $q=4$. Thick, medium, thin lines are for $p=2,3,4$; solid (dashed) lines for bg+1j (bg+2j).}
\end{figure}

The implication of Fig.\ 3 is that the bin size $\delta$ is not an essential variable, and that the normalized asymmetry
\begin{eqnarray}
A_{qp}=\left<D_q^p\right>/\left<S_q^p\right>   \label{9}
\end{eqnarray}
can be a more succinct measure. To see that explicitly we show in Fig.\ 4 the values of $A_{qp}$ vs $\delta^{-1}$ for the 1j and 2j points.  For $q=3$ and 4 the dependences on $p$ and $\delta$ are not sensitive, and the differentiation between the 1j and 2j cases can easily be made, even for $q=2$. It seems that this type of analysis renders more quantitative and distinctive results than 3-particle correlation.

	We have generalized the value of $N_i$ to be Gaussian distributed, and then also the jet multiplicity $m$. We have found no significant effect on the asymmetry measure $A_{qp}$. Since our aim here is only to suggest useful measures, and not their detail numbers, we omit the presentation of the results of the Gaussian-distributed multiplicities, leaving them to be reported elsewhere.

	Restricting our attention now to only the 1j case, we consider the question of how best to reveal the position and shape of a peak on the away side using FM. To that end we need to introduce a cut in $\varphi$. Define
\begin{eqnarray}
f_q^{ ^>_< }(\varphi_c)={1\over M_{ ^>_< }}\sum_{j\in S_{ ^>_< }} n_j(n_j-1)\cdots(n_j-q+1),   \label{10}
\end{eqnarray}
where $S_{ ^>_< }$ is the set of bins in the range $|\varphi|{ ^>_< }\varphi_c$, and $M_{ ^>_< }$ is the number of bins in $S_{ ^>_< }$. The overall FM is then given by $f_q=r_<f_q^< + r_>f_q^>$, where $r_{ ^>_< }=M_{ ^>_< }/M$. The corresponding NFM is
\begin{eqnarray}
F_q^{ ^>_< }(\varphi_c)=f_q^{ ^>_< }(\varphi_c)\left /\left(f_1\right)^q\right. \ ,  \label{11}
\end{eqnarray}
 and the overall NFM is $F_q=r_<F_q^< + r_>F_q^>$.
The relative magnitude of $F_q^>$ vs $F_q^<$, as $\varphi_c$ is varied, provides a measure of the shape of the  $\varphi$ \dis\ of the jet location.  Let us  define 
\begin{eqnarray}
B_q={\left<F_q^< - F_q^>\right>\over \left<F_q^< + F_q^> \right>}. \label{12}
\end{eqnarray}

\begin{figure}[tbph]
\hspace*{-2.8cm}
\vspace*{-.5cm}
\includegraphics[width=0.8\textwidth]{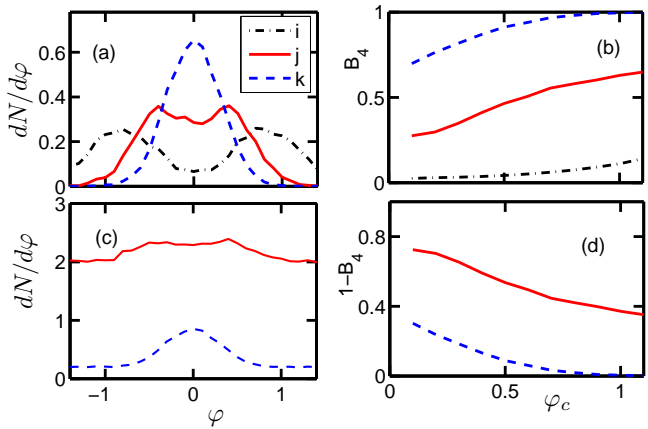}
\vspace*{-1.0cm}
\caption{(a) Event-averaged $\varphi$ distribution for ${\cal N}_{\rm evt}=5000$ with $M=30$. The background contributions are not shown; their levels are [i] 20, [j] 2, [k] 0.2.  The widths and locations of the peaks are: [i] (0.4, 0.8), [j] (0.3, 0.4), [k] (0.3, 0). (b) The corresponding $B_4(\varphi_c)$ distributions. (c) $dN/d\varphi$ for [j] and [k] with bg included. (d) $1-B_4$ for the distributions in (c).}
\end{figure}

	To test its usefulness we consider three cases for which the event-averaged distributions above the background are shown in Fig.\ 5(a); the heights of the background, $dN/d\varphi|_{\rm bg}$, (not shown) are [i] 20, [j] 2, and [k] 0.2. The signal in case [i] is at the 1\% level, and corresponds to very low $p_T^{\rm assoc}$, whereas case [k] has a sharp peak above a low background that can arise from high $p_T^{\rm trig}$ and $p_T^{\rm assoc}$. In Fig.\ 5(b) is shown $B_4(\varphi_c)$ for the three cases, which are well separated. In all cases background alone gives $B_q$ very nearly 0 because either $\left<F_q^<\right>\cong \left<F_q^>\right>$ for $\varphi_c>0.1$ or their sum is much larger than their difference. Note that even in case [i] where the double peaks are small compared to background, we obtain without background subtraction nonzero $B_4(\varphi_c)$, albeit small. In case [j] with $dN/\varphi|_{\rm bg}=2$ the average bin multiplicity $\bar n$ of the background is only 0.2, which is much less than $q\ge 1$. Nevertheless, $f_q^{ ^>_< }$ defined in Eq. (10) does not vanish due to the fluctuations of the bin multiplicities, so $F_q^{ ^>_< }$ exists, though very  large. The event-averaged $\left<F_4^{ ^>_< }\right>$ are between 0 and 4, resulting in $B_4$ to be around 0.5.	
	In case [k] the strong peak at $\varphi=0$ results in $F_q^<$ being large compared to $F_q^>$  except when $\varphi_c$ is small; thus $B_4$ is large, specially at large $\varphi_c$. Where it deviates from 1 is a measure of the narrowness of the peak. $B_q$ for $q=2,3$ give similar results. 
		
		To achieve a better perspective of the mapping between $dN/d\varphi$ and $B_4$, we show in Fig.\ 5(c) the former  for cases [j] and [k] with background included. The corresponding curves for $1-B_4(\varphi_c)$ are shown in (d), in which [j] exhibits the features of both the broad bump in (a) and the high background in (c), while the peak of [k] reflects the same in (a) and (c). These properties are remarkable due to the drastic difference in the nature of the measures displayed. This demonstrates that without background subtraction the FM analysis event-by-event results in clear and quantitative description of the $\Delta\phi$ characteristics.
		
In summary, we have considered the use of factorial moments to analyze the $\Delta\phi$ distribution of particles produced opposite a trigger. The advantage of such a method of analysis is that no explicit background subtraction is necessary. The FM of each event is sensitive to the jet characteristics, while suppressing the effect of statistical fluctuation of the background; it is also insensitive to the smooth variation due to elliptic flow. The asymmetry moments $A_{qp}$ can well separate one-jet and two-jet recoil scenarios, and $B_q$ can give a quantitative description of the single-jet characteristics. Application of this method to the analysis of the RHIC data on jet correlation may provide a common framework to compare results from widely different experimental conditions and various subtraction schemes.

We are grateful to Rene Bellwied and Aihong Tang for helpful comments. This work was supported, in part, by the U.\ S.\ Department of Energy under
Grant No. DE-FG03-96ER40972.


\begin{thebibliography}{99}
\bibitem{ja}
J.\ Adams {\it et al.} (STAR Collaboration), Phys.\ Rev.\ Lett.\ {\bf 95}, 152301 (2005).
 
\bibitem{hb}
H.\ Beusching, nucl-ex/0511044.

\bibitem{dm}
J.\ Adams {\it et al.} (STAR Collaboration), nucl-ex/0604018.

\bibitem{cst}
J.\ Casalderrey-Solana, E.\ V.\ Shuryak, and D.\ Teaney, J.\ Phys.\ Conf.\ Ser.\ {\bf 27}, 22 (2005).

\bibitem{kmw}
V.\ Koch, A.\ Majumder, and X.\ -N.\ Wang, nucl-th/0507063.

\bibitem{id}
I.\ M.\ Dremin, Nucl.\ Phys.\ A {\bf 767}, 233 (2006).

\bibitem{bp}
A.\ Bia\l as and R.\ Peschanski, Nucl.\ Phys.\ B {\bf 273}, 703 (1986); {\bf 308}, 857 (1988).

\bibitem{ddk}
E.\ A.\ De Wolf, I.\ M.\ Dremin, and W.\ Kittel, Phys.\ Rep.\ {\bf 270}, 1 (1996).

\bibitem{bh}
A.\ Bia\l as and R.\ C.\ Hwa, Phys.\ Lett.\ B {\bf 253} 436 (1991).

\bibitem{at}
A.\ Tang (private communication).


\end{thebibliography}
\end{document}